\documentclass[11pt]{article}
\usepackage{amsmath}
\usepackage{amssymb}
\begin{document}
\newcommand{\half}{\frac{1}{2}}
\newcommand\sech{{\rm sech}}
\newcommand{\IZ}[1]{\bar{#1}}
\newcommand{\Deta}{\eta^{\dagger}}
\newcommand\Bpsi{\boldsymbol{\psi}}
\title{
\begin{flushright}
  \small UMP-97/70
\end{flushright}
\vskip1.0cm
\large\bf
Moment Formalisms applied to a solvable Model with a Quantum Phase 
Transition. I. Exponential Moment Methods}
\author
{\large N.S. Witte\footnote{E-mail: {\tt nsw@physics.unimelb.edu.au}}\\
{\it Research Centre for High Energy Physics,}\\
{\it School of Physics, University of Melbourne,}\\
{\it Parkville, Victoria 3052, AUSTRALIA.}\\
{\large R. Shankar}\\
{\it Institute of Mathematical Sciences,}\\
{\it C.I.T. Campus}\\
{\it Taramani, Chennai 600113, INDIA} }
\maketitle
\begin{abstract}
We examine the Ising chain in a transverse field at zero temperature
from the point of view of a family of moment formalisms based upon the
cumulant generating function, where we find 
exact solutions for the generating functions and cumulants at 
arbitrary couplings and hence for both the ordered and disordered
phases of the model.
In a $ t $-expansion analysis, the exact Horn-Weinstein
function $ E(t) $ has cuts along an infinite set of curves in the 
complex $ Jt $-plane which are confined to the left-hand half-plane
$ \Im Jt < -1/4 $ for the phase containing the trial state (disordered), 
but are not so for the other phase (ordered). For finite couplings 
the expansion has a finite radius of convergence. Asymptotic forms 
for this function exhibit a crossover at the critical point,
giving the excited state gap in the ground state sector for the 
disordered phase, and the first excited state gap in the ordered phase.
Convergence of the $ t $-expansion with respect to truncation order
is found in the disordered phase right up to the critical point,
for both the ground state energy and the excited state gap.
However convergence is found in only one of the Connected Moments 
Expansions (CMX), the CMX-LT, and the ground state energy shows
convergence right to the critical point again, although to a 
limited accuracy. 
\end{abstract}
\vskip1.0cm
{\rm PACS: 05.30.-d, 11.15.Tk, 71.10.Pm, 75.10.Jm}
\eject
\section{Motivation}

The purpose of this work is to study the predictions of a family of
non-perturbative moment formalisms devised to treat Field Theoretic and
Many-Body Models with a well known exactly solvable model,
the $ 1-D $ Ising Model in a transverse Field.
This model has a quantum phase transition, and we compare those
predictions of the formalisms with the exact results in the two phases and 
near the critical point of the Model.
These formalisms, the $ t $-Expansion\cite{texp-HW} and the Connected 
Moments Expansion (CMX)\cite{cmx-1st-C}, are related in that they are
centred upon the cumulant generating function, which takes an exponential
form.
These two Expansions have been applied in a number 
of areas ranging from quantum chemistry, lattice gauge theories to 
condensed matter problems and the reader is referred to Witte\cite{texp-W-97}
for references. 
The formalisms take their starting point from a finite sequence of 
low order connected Hamiltonian moments, or cumulants 
$\{ \nu_n : n=1 \ldots \} $, calculated with respect to a chosen trial state,
with the proviso that the overlap with exact ground state is nonzero.
In the $ t $-Expansion, one constructs the cumulant generating function
from its Taylor series expansion about $ t=0 $,
\begin{equation}
   \langle e^{tH} \rangle = 
   \exp\left\{ \sum^{\infty}_{n=1} \nu_n {t^n \over n!} \right\} 
   \equiv e^{F(t)} \ ,
\label{def-F}
\end{equation}
where all moments are calculated with respect to this trial state.
Then the Horn-Weinstein Theorem\cite{texp-HW} states that the ground 
state energy and first excited state gap (within the same sector as
the ground state)\cite{texp-HKW} are given by the following 
limits
\begin{equation}
 \begin{split}
  E_0     & = \lim_{t \to \infty} E(t) \\
  E_1-E_0 & = \lim_{t \to \infty} 
          -{ d^2E(t)/dt^2 \over dE(t)/dt} \ ,
\end{split}
\label{HW-thm}
\end{equation}
where $ E(t) \equiv F'(-t) $ is termed the Horn-Weinstein function.
Other ground state averages can be found by computing the
appropriate generalised connected moments\cite{texp-HW} or higher
excited states\cite{texp-MO} found from certain derivatives of the 
Horn-Weinstein function and performing the same extrapolation. 
This theorem was ``proved'' in Ref.~(\cite{texp-HW}) by assuming
a simple discrete point spectrum for the system, even in the
thermodynamic limit, and these kinds of assumptions continue to
be commonly used without adequate justification.
The question immediately arises of how this extrapolation can be
carried out and several methods\cite{texp-S} are employed. 
These extrapolation methods include Pad\'e Analysis, where
$ E(t) $ is assumed to be representable by a $ [L/M] $ Pad\'e
approximant with $ L=M $ i.e. it has only separable poles and
zeros of finite order in any finite region of the complex
$ t $-plane, and the extrapolation $ t \to \infty $ made directly,
Other methods are D-Pad\'e Analysis, Laplace Method/Resolvent Analysis,
Inversion Analysis, and ``$ E $ of $ F $''/Partition Function Method,
and the reader is referred to Witte\cite{texp-W-97} for the original 
references.

The Connected Moments Expansion (CMX)
is founded upon the assumption of a particular form for the asymptotic 
behaviour of $ E(t) $ as real $ t \to \infty $,
\begin{equation}
   E(t) \sim E_0 + \sum^{n}_{j=1} A_j e^{-b_j t} \ ,
\label{cmx-Et}
\end{equation}
although its variants employ additional and differing assumptions.
The variants give the following explicit forms for the ground state
energy, according to their treatments.
The first method, CMX-HW\cite{cmx-1st-C}, gives the ground state energy 
up the $n$th order as
\begin{equation}
   E_0 = \nu_1-\sum^{n}_{i=1} 
   { S^2_{i,2} \over \prod^{i}_{j=1} S_{j,3} } \ ,
\label{cmxhw-gse}
\end{equation}
where the $ S $-sequence are recursively generated by
\begin{equation}
\begin{split}
   S_{1,k} & = \nu_k \\
   S_{i+1,k} & = S_{i,k}S_{i,k+2}-S^2_{i,k+1} \ .
\end{split}
\label{cmxhw-recur}
\end{equation}
Another method, CMX-SD\cite{cmx-hom-C}, gives the ground state energy up
the $n$th order as
\begin{equation}
   E_0 = \omega_{1,n} \ ,
\label{cmxsd-gse}
\end{equation}
where the $ \omega $-variables are recursively generated by
\begin{equation}
\begin{split}
   \omega_{j,1} & = \nu_j \\
   \omega_{j,k+1} & = 
   \omega_{j,k}-2{ \omega_{j+1,k}\omega_{2,k} \over \omega_{3,k} }
   +\left({ \omega_{2,k} \over \omega_{3,k} } \right)^2\omega_{j+2,k} \ .
\end{split}
\label{cmxsd-recur}
\end{equation}
The last method, the CMX-LT described in Ref.~\cite{cmx-hom-C,cmx-K}, gives
\begin{equation}
   E_0 = \nu_1
   - (\nu_2,\nu_3, \ldots ,\nu_n) \left(
     \begin{array}{cccc}
      \nu_3      &  \nu_4      & \cdots  &  \nu_{n+1} \\
      \nu_4      &  \nu_5      & \cdots  &  \nu_{n+2} \\
      \vdots     &  \vdots     & \ddots  &  \vdots    \\
      \nu_{n+1}  &  \nu_{n+2}  & \cdots  &  \nu_{2n-1}
     \end{array}                  \right)^{-1}
                                  \left(
     \begin{array}{c}
      \nu_2   \\
      \nu_3   \\
      \vdots  \\
      \nu_{n} 
     \end{array}                  \right) \ .
\label{cmxlt-gse}
\end{equation}
There is also another variant, called the AMX\cite{cmx-MZM}, which
is very similar to the CMX-HW method.

Very little has been known of an exact nature concerning these methods.
For example it was not known how the formalisms could be realised in
an exact manner, whether the exact theorems were in fact confirmed
for solvable Models, whether the assumptions made in the inevitable 
extrapolations needed to treat non-integrable Models were
justified, and what factors affected the convergence of the methods.
Recently such questions were begun to be addressed with respect to
another solvable Model, but that Model was studied only at its
critical point - the isotropic XY Model in 1-Dimension\cite{texp-W-97}. 
We extend this analysis here to a closely related Model, within the
same universality class, and reexamine these questions with regard
to a phase transition.
We firstly analytically solve our Model from the moment point of view,
then investigate the validity of the theorems using the exact
Horn-Weinstein function, 
following this we reveal the analytic character of this function in
the complex $ t $-plane,
and in the truncated expressions using a set of
finite low order cumulants examine the convergence of ground state 
spectral quantities with respect to the order of truncation.
In a companion work to this paper, paper II\cite{alm-itf-W-98}, we
continue analysis of this model from the point of view of the other
family of moment methods - the geometrical family, which includes the
Lanczos Method.

\eject

\section{The ITF Model}

The model we investigate is the Ising model in a transverse field
in one dimension, which exhibits a order-disorder quantum phase
transition at zero temperature\cite{itf-P,itf-CDS}.
The Hamiltonian for this Model is taken to be the following
\begin{equation}
 H = -J\sum^{N}_{n=1} \sigma^{z}_{n}\sigma^{z}_{n+1}
     +B\sum^{N}_{n=1} \sigma^{x}_{n} \ ,
\label{itf-ham}
\end{equation}
which differs slightly in conventions with previous use.
We define a coupling constant ratio $ x = B/J $. 
In what follows the label $ n $ indexes the sites and periodic boundary 
conditions $ \sigma^{a}_{N+1} = \sigma^{a}_{1} $ are applied.
This model has a 
zero-temperature quantum phase transition at $ B = J $ or $ x = 1 $, 
which separates an ordered phase $ x < 1 $ whereby the z-component of all 
the spins are aligned either up or down from the disordered phase 
$ x > 1 $ induced by the tunnelling term. This model is also self-dual, 
in that $ H(x) \to xH(1/x) $ leads to the same model, and consequently for
the ground state properties. The exact ground state is given by
\begin{equation}
  \epsilon_0/J =
  - {2\over \pi}|1+x| 
    E(\sqrt{\scriptstyle 4|x| \over \scriptstyle (1+x)^2}) \ ,
\label{itf-gse}
\end{equation}
where $ E(k) $ is the complete Elliptic function of the second kind, 
according to the conventions of Gradshteyn and Ryzhik\cite{integral-GR}.
The first excited state gap has the exact form
\begin{equation}
   \Delta E = 2J|1-x| \ ,
\label{itf-gap}
\end{equation}
which vanishes at the transition point.
For the full details of the solution to this model and discussion of its
properties we refer the reader to the works\cite{itf-P,itf-K,itf-CDS}.

In the first step to an explicit form for the cumulant generating function
one makes a transformation from the ordered spin operators to the 
disordered spin operators
\begin{equation}
  \mu^{1}_{n} = \prod^{n-1}_{m=1} \sigma^{x}_{m} \ , \quad
  \mu^{2}_{n} = \sigma^{z}_{n}\sigma^{z}_{n-1} \ , \quad
 i\mu^{3}_{n} = \mu^{1}_{n}\mu^{2}_{n} \ ,
\label{od-xfm}
\end{equation}
and then the Wigner-Jordan transformation to a pure Fermionic representation
\begin{equation}
  c_{n}     = \mu^{1}_{n} \sigma^{-}_{n} \ , \quad
  c^{+}_{n} = \sigma^{+}_{n} \mu^{1}_{n} \ .
\label{wj-xfm}
\end{equation}
In doing so we have neglected the boundary terms arising from the exact
mapping of the spin or a-cyclic problem, and will only consider the
c-cyclic problem $ c_{N+1} = c_1 $. 
This followed by a Fourier transformation of the Fermionic operators
\begin{equation}
  c_{k} = {1 \over \sqrt{N}} \sum^{N}_{n=1} e^{ikn} c_{n} \ ,
\label{f-xfm}
\end{equation}
where the momenta are $ k = 2\pi l/N $ and the integers
$ l = -N/2, \ldots, N/2-1 $ for $ N $ even, whereas
$ l = -(N-1)/2, \ldots, (N-1)/2 $ for $ N $ odd.
Proceeding in this way the Hamiltonian becomes partially decoupled as
\begin{equation}
  H = \sum_{k>0} h_{k}   
    = \sum_{k>0} \left\{  
                 2iJ\sin k [ c^{+}_{k}c^{+}_{-k}\!+\!c_{k}c_{-k} ]
         +2(J\cos k\!+\!B) [ c^{+}_{k}c_{k}\!+\!c^{+}_{-k}c_{-k}\!-\!1 ]
                 \right\} \ .
\label{mom-H}
\end{equation}
For each $ k $ there is a 4-dimensional Hilbert space
$ |00\rangle, |01\rangle, |10\rangle, |11\rangle $
but because 
$ c^{+}_{k}c^{+}_{-k}, c_{k}c_{-k} $ only connect
$ |00\rangle, |11\rangle $ and 
$ c^{+}_{k}c_{k}, c^{+}_{-k}c_{-k} $ are diagonal
one only needs the 2-dimensional BCS reduced subspace
$ |00\rangle, |11\rangle $.
Thus we have a pseudo-spin problem
$ |\Downarrow\rangle = |00\rangle, |\Uparrow\rangle = |11\rangle $
with pseudo-spin operators
\begin{equation}
  \tau^{+}_{k} = c^{+}_{k}c^{+}_{-k} \ , \quad
  \tau^{-}_{k} = c_{-k}c_{k} \ , \quad
  \tau^{3}_{k} = c^{+}_{k}c_{k}\!+\!c^{+}_{-k}c_{-k}\!-\!1 \ ,
\label{p-spin}
\end{equation}
and thus
$ h_{k} = -2J\sin k \tau^{2}_{k} + 2(J\cos k\!+\!B) \tau^{3}_{k} $.

As an obvious first choice for the trial state, but not the only choice, 
we take the exact eigenstate as $ x \to \infty $, that is the 
ferromagnetically aligned state $ |{\cal F} \rangle $ polarised along the
$ x $-axis
$ | \Downarrow \rangle =
  | \downarrow \downarrow \downarrow \ldots 
    \downarrow \downarrow \downarrow \rangle_{x} $,
which is represented in the fermionic $c-$picture by
$ | 0 0 0 \ldots 0 0 0 \rangle $.
Another equally obvious choice for the trial state would be the exact 
eigenstate for $ x = 0 $, and this we denote $ |{\cal F}^{\#} \rangle $, and 
is the fully polarised state in the $z$-direction. This implies that 
we only need the matrix elements with respect to
\begin{equation}
  \langle \Downarrow_{k}| e^{th_{k}} |\Downarrow_{k}\rangle =
  \cosh(tE_{k}) - {2(J\cos k\!+\!B) \over E_{k}}\sinh(tE_{k}) \ ,
\label{mgf-factor}
\end{equation}
where the quasi-particle energy is
\begin{equation}
  E^2_{k} = 4J^2+4B^2+8JB\cos k \ .
\label{dqp-energy}
\end{equation}
The moment generating function is then quite simply given by
\begin{equation}
\begin{split}
  \langle e^{tH} \rangle & =
    \prod_{k>0} \left\{ 
  \cosh(tE_{k}) - {2(J\cos k\!+\!B) \over E_{k}}\sinh(tE_{k})
                \right\} \\
                & \xrightarrow[N \to \infty]{} 
    \exp \left\{ {N \over 2\pi} 
         \int^{\pi}_{0} dk\, 
         \ln \left[
  \cosh(tE_{k}) - {2(J\cos k\!+\!B) \over E_{k}}\sinh(tE_{k})
         \right]
         \right\} \ .
\end{split}
\label{mgf}
\end{equation}
To simplify our discussion we move to dimensionless quantities and
the dimensionless quasi-particle energy squared $ E_{k} = 2J\epsilon_{k} $
is then
\begin{equation}
  \epsilon^2_{k} = 1+x^2+2x\cos k \ .
\label{qp-energy}
\end{equation}
It should be noted that it was not necessary to go any further and fully
diagonalise the problem in order to find the cumulant generating 
function.

However continuing with the diagonalisation process one employs a Bogoliubov 
transformation to quasi-particle operators
$ \eta_{q},\eta^{+}_{q} $ 
\begin{equation}
\begin{split}
  c_{q} & = u_{q} \eta_{q} - i v_{q} \eta^{+}_{-q} \ , \quad
   c_{-q} = u_{q} \eta_{-q} + i v_{q} \eta^{+}_{q} \ , \\
  c^{+}_{q} & = u_{q} \eta^{+}_{q} + i v_{q} \eta_{-q} \ , \quad
   c^{+}_{-q} = u_{q} \eta^{+}_{-q} - i v_{q} \eta_{q} \ , \\
\end{split} 
\label{b-xfm}
\end{equation}
with transformation elements
\begin{equation}
  u_{q} =
  \left\{ { \epsilon_{q}\!+\!x\!+\!\cos q \over 2\epsilon_{q} }
  \right\}^{1/2} \ , \qquad
  v_{q} =
  \left\{ { \epsilon_{q}\!-\!x\!-\!\cos q \over 2\epsilon_{q} }
  \right\}^{1/2} \ .
\label{b-coeff}
\end{equation}
The trial state can now be expressed in terms of quasi-particle operators
and the exact ground state by
\begin{equation}
  |{\cal F}\rangle = \prod_{q>0} {
  \sqrt{ \vphantom{|}\epsilon_{q}\!+\!x\!+\!\cos q }
  + i\sqrt{ \vphantom{|}\epsilon_{q}\!-\!x\!-\!\cos q }
            \,\eta^{+}_{q}\eta^{+}_{-q}
     \over \sqrt{ 2\epsilon_{q} } } |0\rangle \ ,
\label{trial-gs}
\end{equation}
and the overlap squared simply reduces to
\begin{equation}
 \ln |\langle {\cal F}|0 \rangle |^2 =  
 {N \over 2\pi} \int^{\pi}_{0} dq\, 
     \ln \left({ \epsilon_{q}\!+\!x\!+\!\cos q \over 2\epsilon_{q} }
         \right) \ .
\label{overlap}
\end{equation}
As $ x \to \infty $ then
$ |\langle {\cal F}|0 \rangle|^2 \to 1 $, as it must, while for 
$ x = 1 $ we have
\begin{equation}
  \ln |\langle {\cal F}|0 \rangle|^2 = (-\ln 2 + 2G/\pi )N
\end{equation}
where $ G $ is Catalan's constant, and finally at $ x = 0 $, 
$ |\langle {\cal F}|0 \rangle|^2 = 2^{-N} $.
Clearly the logarithm of the overlap squared is simply proportional
to $ N $ with a negative coefficient which is zero at $ x = \infty $
and decreases monotonically to $ -\ln 2 $ at $ x = 0 $.

The lowest excited state can be found from the moment problem
associated with a trial state with the correct quantum numbers,
and the obvious candidate is
\begin{equation}
  | {\cal E} \rangle = {1 \over \sqrt{N}} \sum^{N}_{n=1} \sigma^{+}_{n} 
  | {\cal F} \rangle
\label{trial-E}
\end{equation}
which can be shown to be proportional to
$ c^{+}_{k=\pi} | {\cal F} \rangle $.
The relevant moment generating function is then simply
\begin{equation}
  R(t) \equiv 
  {\langle {\cal E} | e^{-tH} | {\cal E} \rangle \over
   \langle {\cal F} | e^{-tH} | {\cal F} \rangle}
  = e^{-2t|J-B|} 
\label{cgf-E}
\end{equation}
because the quasiparticles are non-interacting in this model.

\eject

\section{The Analytic Character of $ E(t) $}

The analytic character of the Horn-Weinstein function $ E(t) $ in the
complex $ t $-plane is of fundamental importance in devising any
extrapolation strategy for the $ t $-expansion, as knowledge even of
a qualitative nature or the lack of it is the greatest source of errors
in extrapolation.
From the moment generating function the exact Horn-Weinstein function 
$ E(t) $ is given by
\begin{equation}
   E(t) = -{NJ \over \pi}
   \int^{\pi}_{0}dk\, \epsilon_{k}
   { {\displaystyle x\!+\!\cos k \over \displaystyle \epsilon_{k}}
     + \tanh(2Jt\epsilon_{k})
     \over
     1 + {\displaystyle x\!+\!\cos k \over \displaystyle \epsilon_{k}}
         \tanh(2Jt\epsilon_{k}) } \ .
\label{HW-E}
\end{equation}

Before treating the case of general $ x $, the two extreme cases of
$ x \to 0,\infty $ are taken first, which lead to simple results.
The limiting form of $ E(t) $ when $ x \to \infty $ is, of course,
completely trivial, namely $ E(t) \to -NB $.
As will become apparent later however the way this limit is approached
exhibits some complex character. 
In the case when $ x \to 0 $ some care in the limit needs to be taken
and the following result holds
\begin{equation}
  E(t) =
  \begin{cases}
    -NJ\tanh(Jt)
    \qquad & 
    (4K\!-\!1)\frac{\pi}{4} < \Im(Jt) < (4K\!+\!1)\frac{\pi}{4} \\
    -NJ\coth(Jt)
    \qquad & 
    (4K\!-\!3)\frac{\pi}{4} < \Im(Jt) < (4K\!-\!1)\frac{\pi}{4} \\
  \end{cases}
  \quad K \in {\mathbb Z} \ .
\label{HW-E-0}
\end{equation}
The analytic form this function has is one with an infinite number of
Riemann sheets bounded by cuts at $ Jt = (2K\!+\!1)\frac{\pi}{4}i $ for all
$ K \in {\mathbb Z} $.
There are no divergences in this function anywhere, while it has 
discontinuities on the branch lines defined above.
Correspondingly the radius of convergence of the expansion about 
$ t = 0 $ is not $ \pi/2 $, as might be expected from considerations
of the growth in the cumulants, but is $ \pi/4 $.

\begin{figure}[htb]
 \vskip 23.2cm
 \includegraphics{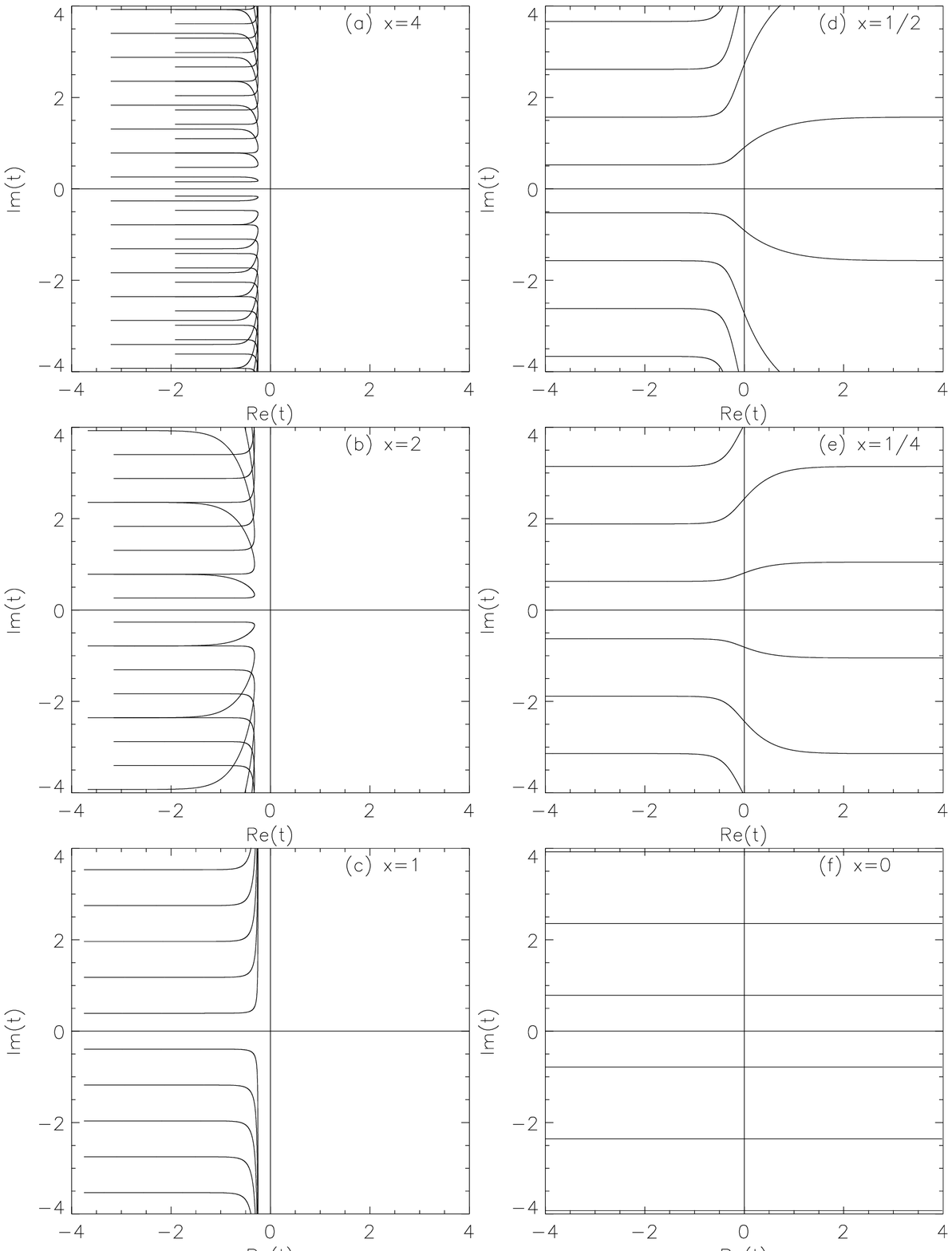}
 \caption
   [The singular curves of $ E(t) $ in the complex $ Jt$-plane for
    various couplings, (a) $x=4$, (b) $x=2$, (c) $x=1$, (d) $x=1/2$,
    (e) $x=1/4$, and (f) $x=0$. ]{}
\label{fig-sing-curve}
\end{figure}

\begin{figure}[htb]
 \vskip 22.7cm
 \includegraphics{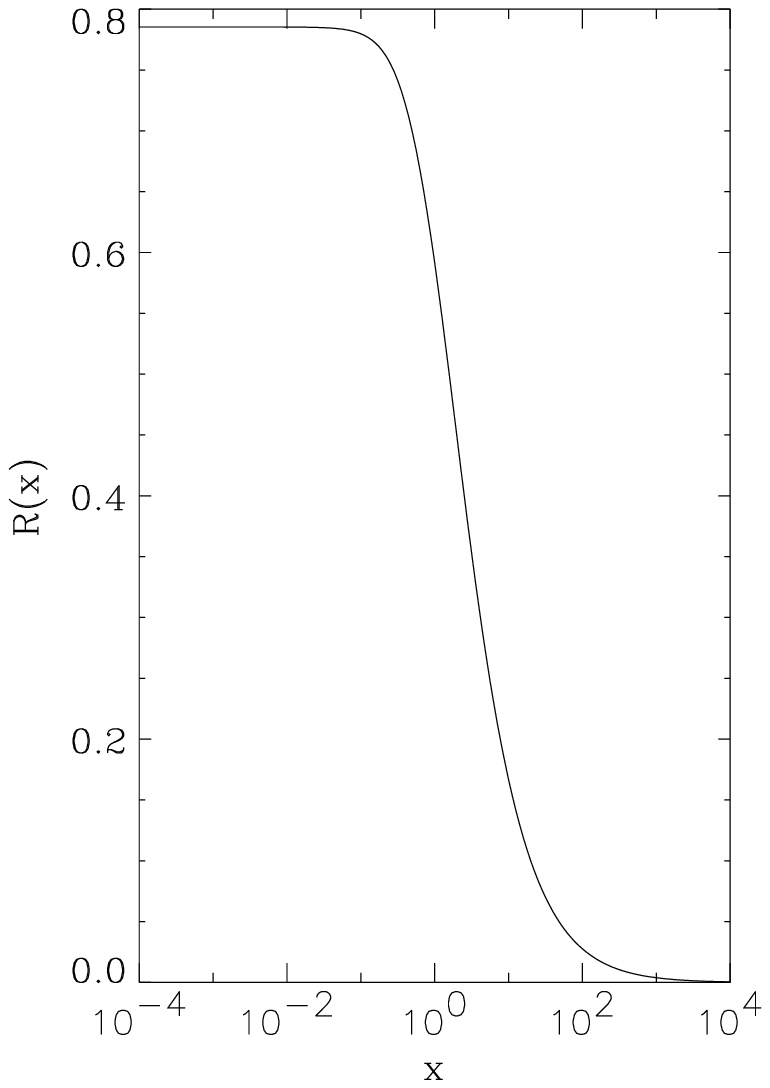}
 \caption
   [The radius of convergence for the Taylor expansion of $ E(t) $ 
    about $ t=0 $ versus coupling $ x $.]{}
\label{radius}
\end{figure}

The singularity structure of $ E(t) $ for general $ x $ is most interesting
as it is a singular integral, and has discontinuities across a
denumerable set of open, infinite curves in the complex $ t $-plane. 
This set of singular curves, where the denominator of the integrand of 
Eq.~(\ref{HW-E}) vanishes, is defined by
\begin{equation}
\begin{split}
  \Im\, Jt & = (K+1/2){\pi\over 2e} \qquad K \in {\mathbb Z}
  \\
  \Re\, Jt & = {1\over 2e} \tanh^{-1}\left\{
               {1\over 2x}\left({1-x^2\over e}-e\right) \right\} \ ,
\end{split}
\label{sing-curve-x}
\end{equation}
and parametrised by a real $ e $ with $ |x-1| \leq e \leq x+1 $ when 
$ x \neq 1 $, and by
\begin{equation}
\begin{split}
  \Im\, Jt & = (K+1/2){\pi\over 4\xi} \qquad K \in {\mathbb Z}
  \\
  \Re\, Jt & = -{1\over 4\xi} \tanh^{-1}(\xi) \ ,
\end{split}
\label{sing-curve-1}
\end{equation}
when $ x=1 $, with $ 0 < \xi < 1 $.
This family of curves in the complex $ Jt $-plane can be simply described
as the different Riemann sheets of the following transcendental equation
\begin{equation}
    \exp\left( 2Jt \sqrt{x+z}\sqrt{x+1/z} \right) =
  { \sqrt{x+z}-\sqrt{x+1/z} \over \sqrt{x+z}+\sqrt{x+1/z} } \ ,
\label{riemann-sheets}
\end{equation}
with an arbitrary complex parameter $ |z| = 1 $, and
where the only choices of the branch or sheet arise from the logarithm
and not the square-roots.
In Figure~(\ref{fig-sing-curve}) these curves are shown for a sequence of
$x$-values from $4$ to $0$, and a remarkable change occurs at the
transition point. For $ x > 1 $ the singular curves are confined to
the left-half of the complex $Jt$-plane, to the left of
\begin{equation}
  \Re\, Jt_0 = -{1 \over 2\sqrt{x^2-1}}\cosh^{-1}(x) \ ,
\end{equation}
with both asymptotes of the curves bent back parallel to the real
axis. At $ x=1 $ the asymptotes furthest from the real axis have opened
up and become parallel to the imaginary axis. They are still confined to
$ \Re\, Jt < -1/4 $.
However once $ x<1 $ these asymptotes invade the right hand side of
the $Jt$-plane, becoming parallel to the real axis as
$ \Re\, Jt \to +\infty $, although tending towards a different 
$ \Im\, Jt $ value from that of the left-hand asymptote. Finally at
$ x=0 $ the singular curves are straight and parallel to the real axis.
Thus we see that the big shift from one phase to the other is marked
by the loss of analyticity in the right-hand side of the complex
$Jt$-plane. Expressed in another way one can perform a Wick rotation from
the positive real axis through $ \pm \pi/2 $ without encountering any
singularities for $ x \geq 1 $, but for $ x < 1 $ this is not possible.
With this information we can find the radius of convergence of the
Taylor series expansion about $ t=0 $, from the distance of the 
singular curves for $ K=0,-1 $ from the origin.
If we denote this by $ R(x) $, we have
\begin{equation}
  R(x) = {1\over 4e_0}\sqrt{ \pi^2 + 
         \ln^2\left|{ 1-(e_0\!-\!x)^2 \over (e_0\!+\!x)^2-1 }\right| }\ ,
\label{rad-conv}
\end{equation}
where the parameter $ e_0 $ is given the solution of
\begin{equation}
   \ln^2\left|{ 1-(e_0\!-\!x)^2 \over (e_0\!+\!x)^2-1 }\right|
   +{4xe_0(e^2_0-x^2+1) \over [(x\!+\!1)^2-e^2_0][e^2_0-(x\!-\!1)^2]}
   \ln\left|{ 1-(e_0\!-\!x)^2 \over (e_0\!+\!x)^2-1 }\right|
   + \pi^2 = 0 \ .
\end{equation}
The radius of convergence approaches $ \pi/4 $ for small $ x$-values
and decreases monotonically as $ x $ is increased, finally taking the
form of $ \ln(2x)/2x $ for large $ x $. This behaviour is shown in
Figure~(\ref{radius}).
However the actual size of the radius of convergence is not important in
this method, although it should be non-zero, but rather whether the 
right half of the complex $ Jt $-plane is free of singularities.

As a final step in describing the analytic nature of this function we
give the following result for the discontinuity in it across the
singular curves,
\begin{equation}
\begin{split}
  \Delta E(Jt) =  &
 -{8i NJ e^3 \over  \sqrt{(x\!+\!1)^2-e^2}\sqrt{e^2-(x\!-\!1)^2}}
                  \\
  & \qquad \times  \left\{
  {4x e(e^2\!+\!1\!-\!x^2) \over  [(x\!+\!1)^2-e^2][e^2-(x\!-\!1)^2]}
     + \ln\left|{ 1-(e\!-\!x)^2 \over (e\!+\!x)^2-1 }\right|
       +i\pi(2K+1) \right\}^{-1} \ .
\end{split}
\label{discontinuity}
\end{equation}
Asymptotically along either arm of the curves 
$ e \to |1-x|,1+x $ this jump vanishes for all $ x $ values, while there
are single, local but finite enhancements at some intermediate value of $ e $
(a slightly different point for the real and imaginary parts). This 
appears as a sharp enhancement for all curves which occurs close to the
imaginary axis, but the jump here is not divergent (it is not possible for
the denominator of Eq.\ref{discontinuity} to vanish).
Thus all of the singular nature resides in discontinuity of the function
in passing from one Riemann sheet to another.

\eject

\section{The Asymptotic Forms of $ E(t) $}

Of central importance in understanding the $ t $-expansion theorems in
our Model is the asymptotic form for $ E(t) $ as $ t \to \infty $ in
a sector which contains the positive real axis.
There several reasons for this. Firstly any changes in the asymptotic
behaviour with coupling reflects changes in the analytic character of
$ E(t) $ with coupling also, secondly it has a bearing on extrapolation
strategies for the $ t $-expansion, and lastly it gives the precise
forms that should be assumed for Connected Moment Expansions.
The explicit form for the Horn-Weinstein function satisfies the ground 
state Horn-Weinstein theorem in that 
$ \lim_{t \to \infty} E(t) \to \epsilon_0 $ where the ground state energy 
is given by Eq.~(\ref{itf-gse}).
To find higher order asymptotic terms we have to consider the cases 
$ x>1 $, $ x=1 $, and $ x<1 $ separately as we will find a crossover in 
the asymptotic behaviour at the transition point. In the first case, 
$ x>1 $, one can make a simple expansion of the denominator of the 
integrand, integrate term-by-term, and apply a variant of a lemma of 
Erd\'elyi\cite{asymptotic-E} to arrive at
\begin{equation}
\begin{split}
  E(t)-E_{\infty} 
  & \sim
  -{2NJ\over \pi} \sum_{k=1}^{\infty} (-)^k
   \left\{ (x\!-\!1)^2 e^{-4Jtk(x\!-\!1)} \sum^{\infty}_{n=k}
           {\Gamma(n\!+\!1/2)\over n![8Jtkx(x\!-\!1)]^{n+1/2}} U^0_{nk}(x)
   \right.
  \\
  & \phantom{\sim -{2NJ\over \pi} \sum_{k=1}^{\infty} (-1)^k \Biggl\{}
   \left. \pm i(x\!+\!1)^2 e^{-4Jtk(x\!+\!1)} \sum^{\infty}_{n=k}
           {\Gamma(n\!+\!1/2)\over n![8Jtkx(x\!+\!1)]^{n+1/2}} U^1_{nk}(x)
   \right\}
  \\
  & \sim {NJ\over \pi^{1/2}} {(x\!-\!1)^{1/2}\over [8Jtx]^{3/2}}
         e^{-4Jt(x\!-\!1)}
\end{split}
\label{bigx-asympt}
\end{equation}
for $ |{\rm Arg}\,Jt| \leq \pi/2 $. This expression is not valid at $ x = 1 $.
The $ U^{0,1}_{nk}(x) $ are polynomials in $ x $ with integer coefficients, 
defined by the generating functions
\begin{equation}
\begin{split}
  \sum^{\infty}_{n=0}{z^n\over n!}U^{0}_{nk}(x) & =
  z^k(1+2xz)^2[1-x(x\!-\!1)z]^{k-1/2}[1+(x\!-\!1)z]^{-k-1/2}[1+xz]^{-k-1/2}
  \ , \\
  \sum^{\infty}_{n=0}{z^n\over n!}U^{1}_{nk}(x) & =
  z^k(1+2xz)^2[1+x(x\!+\!1)z]^{k-1/2}[1+(x\!+\!1)z]^{-k-1/2}[1+xz]^{-k-1/2}
  \ .
\end{split}
\label{p0-defn}
\end{equation}
Clearly, in this phase, the asymptotic form has a dominant exponential 
decay controlled by the first excited state gap $ 4J(x-1) $ in the
ground state sector together with the
algebraic factors $ t^{-n-1/2} $ for $ n \geq 1 $. There are also
subdominant terms with an exponential decay controlled by the gap
at the top of the spectrum $ 4J(x+1) $,
which are discontinuous as $ {\rm Arg}\,Jt $ 
changes sign as a manifestation of the Stoke's phenomena.
The Stoke's line is $ {\rm Arg}\,Jt = 0 $, on which the first asymptotic
series has maximal exponential dominance over the second series, or its
associated series. The Anti-Stoke's line are $ {\rm Arg}\,Jt = \pm\pi/2 $
where the asymptotic and its associated series are of the same order.
For the purposes of the Horn-Weinstein theorem
within this sector of the $ t $-plane the dominant term in the asymptotic
expression above is the exponential term with a unit multiple of the
gap $ 4J(x-1) $, then the higher algebraic terms combined with this,
while all other exponential terms are subdominant. The relative ordering
of these latter terms can vary with coupling but has no physical significance
for the Horn-Weinstein theorems.
It can also be shown that the excited state Horn-Weinstein theorem
when applied to the above function $ E(t) $ yields
\begin{equation}
   \lim_{t \to \infty} {-E''(t) \over E'(t)}
   = 4J(x\!-\!1) \ .
\label{bigx-HW-gap}
\end{equation}
which arises from the excitation of two noninteracting quasi-particles 
near momenta $ k = \pi + {\rm O}(1/N) $.

Exactly at the critical point the leading exponential factors vanish
leaving the dominant series having only algebraic decay of the form,
\begin{equation}
\begin{split}
  E(t)-E_{\infty}
  & \sim
  -{8NJ\over \pi} \sum_{k=1}^{\infty} (-)^k
   \left\{ \sum_{n=1} [8Jtk]^{-n-1} V_{nk}
   \right.
  \\
  & \phantom{\sim -{8NJ\over \pi} \sum_{k=1}^{\infty} (-)^k \Biggl\{}
   \left.  +e^{-8Jtk} \sum_{n=k} 
         {\Gamma(n\!+\!1/2)\Gamma(1/2\!-\!k)(n\!-\!2k\!+\!1/2)
          \over (n\!-\!k)!\Gamma(3/2\!-\!n)[16Jtk]^{n+1/2}}
   \right\}
  \\
  & \sim {8NJ \over \pi} \left\{
    {\pi^2\over 12}[8Jt]^{-2} - {\pi^2\over 3}[8Jt]^{-3}
       + {7\pi^4\over 120}[8Jt]^{-4} \ldots \right\}
\end{split}
\label{critical-asympt}
\end{equation}
with $ V_{nk} $ defined by the generating function
\begin{equation}
  \sum^{\infty}_{n=0}{z^n\over n!}V_{nk}(x) =
  z(1-z)^{k-1/2}(1+z)^{-k-1/2}
  \ .
\label{G-defn}
\end{equation}
This expression is still valid for $ |{\rm Arg}\,Jt| < \pi/2 $, as before.
The prediction of the Horn-Weinstein theorems from this case would be that
the first excited state gap would vanish, and correctly so.

However when $ x < 1 $ something interesting and different happens from that
found for $ x \geq 1 $.
Firstly it appears that an asymptotic expansion in the above sense is no
longer possible, in that such an expansion should be valid within a finite 
sector containing the positive real axis. From the previous section we
know that the Horn-Weinstein function is only analytic within a sector of
zero extent about the positive real axis, and any asymptotic expansion can
only be valid within this. Furthermore it is clear from the defining 
$ t $-integrals, Eq.\ref{HW-E}, that for $ x<1 $ the integrand has an inverse
square-root singularity at the lower limit of the integral in contrast
to the case of $ x>1 $ which has a square-root branch point at both
limits of integration. 
It is the algebraic singularities at the end-points
of the integration $ t $-interval which control the exponential factors in 
the asymptotic expansions. Instead of a full asymptotic expansion we
recover just the leading order term,
\begin{equation}
  E(t)-E_{\infty} 
  \sim 2NJ (1\!-\!x)^{2} e^{-2Jt(1\!-\!x)} \ ,
\label{smallx-asympt}
\end{equation}
which implies that the excited state gap is now
\begin{equation}
   \lim_{t \to \infty} {-E''(t) \over E'(t)}
   = 2J(1\!-\!x) \ .
\label{smallx-HW-gap}
\end{equation}
In this phase the Horn-Weinstein excited state theorem gives the lowest
excited state, rather than the next one lying in the ground state
sector, because the trial state 
appropriate for $ x > 1 $ has a finite projection onto the first excited
state for $ x < 1 $ (it is exactly orthogonal to the first excited state
for $ x > 1 $, $ |{\cal E}\rangle $), and it gives the smaller of the 
two gaps.

The lowest excited state gap is found from the generating function
given in Eq.~(\ref{cgf-E}) and it is trivially and exactly true that
\begin{equation}
   \lim_{t \to \infty} {-R'(t) \over R(t)}
   = 2J|1-x| \ ,
\end{equation}
for all $ x $ which is just the energy of the lowest quasiparticle energy 
at $ k = \pi $.

\eject

\section{The Cumulants}

In the application of the t-expansion to non-integrable models, one
computes cumulants with respect to the trial state up to some finite
order and uses the truncated cumulant generating function in the
ensuing analysis. Because our convergence studies will do just this,
we need to compute the cumulants. Furthermore issues concerning the
radius of convergence of the Taylor expansion of $ E(t) $ about the
origin will arise in any discussion of the analytic character of this
function.
It is sometimes convenient to discuss normalised cumulants $ c_n/J^{n} $
which are polynomials in $ x $ and we will often use the same symbol 
to denote these as there can be no risk of confusion.
The first few normalised cumulants are then
\begin{equation}
 \begin{split}
  c_1  & = -x \\
  c_2  & =  1 \\
  c_3  & = 4x \\
  c_4  & = 16x^2-2 \\
  c_5  & = 64x^3-48x \\
  c_6  & = 256x^4-672x^2+16 \ .
 \end{split}
\end{equation}
The general form for the cumulant coefficients has an integral 
representation
\begin{equation}
  c_{n+1} = { (2J)^{n+1} \over 2\pi }
  \int^{+\infty}_{-\infty} {dy \over \sqrt{\cosh^2 y\!-\!x^2}}
  {d^n \over dy^n}\tanh y 
  \left[{ \sqrt{\cosh^2 y\!-\!x^2} - x\sinh y \over \cosh y }\right]^{n+2}\ ,
\label{cumulant}
\end{equation}
found using the change of variables
\begin{equation}
  \tanh y = -{ \cos k\!+\!x \over \epsilon_{k} } \ .
\end{equation}
The cumulant is a polynomial in $ x $ despite appearances which is a
consequence of choosing the trial state to be an exact eigenstate of the
second term in the Hamiltonian, Eq.~(\ref{itf-ham}).

The explicit forms for these cumulants are then
\begin{equation}
\begin{split}
  {c_{2n+1}\over (2J)^{2n+1}} & =
  -\sum^{n}_{p=0} 
       {(2p)!\over 2^{p+1}p!}{(n\!-\!p)!\over (n\!+\!1\!+\!p)!}
   {|1\!-\!x^2|^{n+1}\over (2x)^{p+1}} A^n_p
   \left\{ {\rm sgn}(x^2\!-\!1)(n\!+\!1\!+\!p)
           P^{p}_{n}\left(\left|{1\!+\!x^2\over 1\!-\!x^2}\right|\right)
   \right.
   \\
   & \left. \phantom{ =
  -\sum^{n}_{p=0} 
       {(2p)!\over 2^{p+1}p!}{(n\!-\!p)!\over (n\!+\!1\!+\!p)!}
   {|1\!-\!x^2|^{n+1}\over (2x)^{p+1}} A^n_p
   \Biggl\{ }
         + (n\!+\!1\!-\!p)
           P^{p}_{n+1}\left(\left|{1\!+\!x^2\over 1\!-\!x^2}\right|\right)
   \right\} \ ,
   \\
  {c_{2n+2}\over (2J)^{2n+2}} & =
   \sum^{n+1}_{p=0} 
       {(2p)!\over 2^{p+1}p!}{(n\!+\!1\!-\!p)!\over (n\!+\!1\!+\!p)!}
   {|1\!-\!x^2|^{n+1}\over (2x)^{p}} B^n_p
           P^{p}_{n+1}\left(\left|{1\!+\!x^2\over 1\!-\!x^2}\right|\right)
   \ ,
\end{split}
\label{cum-poly}
\end{equation}
where the $ P^m_n(z) $ are the associated Legendre Polynomials and the 
$ A^n_p, B^n_p $ are integer coefficients defined from the higher 
derivatives of $ \tanh(z) $,
\begin{equation}
\begin{split}
   {d^{2n}\over dz^{2n}} \tanh(z) & =
   \tanh(z)\sum^{n}_{p=0} A^{n}_{p} \sech^{2p}(z) \ ,
   \\
   {d^{2n+1}\over dz^{2n+1}} \tanh(z) & =
   \sum^{n+1}_{p=0} B^{n}_{p} \sech^{2p}(z) \ .
\end{split}
\end{equation}

One of the simplest explicit forms of these coefficients is one
in terms of the Stirling numbers of the second kind,
$ S(n,k) $, in the following way
\begin{equation}
  A^{n}_{p} = (-)^p 2^{2(n-p)} \sum^{2n}_{k=2p} (-)^k k!
  {k-p-1 \choose p-1} S(2n,k) \ ,
\label{A-sum}
\end{equation}
and the other coefficients are given by 
$ A^{n+1}_{p} = -2p B^{n}_{p} $.

The cumulants with respect to the trial state $ | {\cal F}^{\#} \rangle $
are very simply given by
\begin{equation}
  c^{\#}_n(J,B) = c_n(B,J)
\end{equation}
and these normalised cumulants are then related by
\begin{equation}
  c^{\#}_n(x) = x^{n}c_n(1/x) \ ,
\end{equation}
i.e. the reciprocal polynomials.
The cumulants simplify for the three special cases
$ x = \infty,1,0 $.
In the case $ x \to \infty $, where the $ J \to 0 $ limit is taken and 
$ B $ fixed, then only the first cumulant is non-zero, $ c_1 = -B $, as 
would be expected. At the transition point $ x = 1 $ then
\begin{equation}
  c_{n+1} = (-4J)^{n+1} {1 \over \pi}
  \int^{\infty}_{0} dy\, \sech y \tanh^{n+1}\!y
  {d^n \over dy^n}\tanh y \ ,
\label{cumulant-1}
\end{equation}
which does not appear to be expressible in terms of any known
mathematical constants, and forms a new integer sequence\cite{njas-97}.
At $ x = 0 $ the odd cumulants vanish while the even ones are 
given by
\begin{equation}
  c_{2k} = (2J)^{2k}{(2^{2k}-1) \over 2k} B_{2k} \ .
\label{cumulant-0}
\end{equation}
where the $ B_{2k} $ are the Bernoulli numbers.

\eject

\section{Convergence Studies}

Of great practical interest is the question of convergence of the
the various formalisms arising out of the $ t $-expansion and how
they change crossing the transition point,
and here we examine all the dependences of the errors on the order of
truncation for the ground state energy and the excited state gap
as functions of coupling.
In the results reported here all quantities were calculated using
exact rational arithmetic including cumulants, truncated generating
functions, Pad\'e approximants, matrix inner products, solutions of
the CMX recurrence relations, and CMX energies using the computer
algebra system Maple V. Conversion to approximate floating point numbers
(number of digits was 32) was only made for the final answer, so
loss of precision or roundoff issues do not arise here. All data files
and Maple source code are available upon request to one of the 
authors (NSW).

\begin{figure}[htb]
 \vskip 22.7cm
 \includegraphics{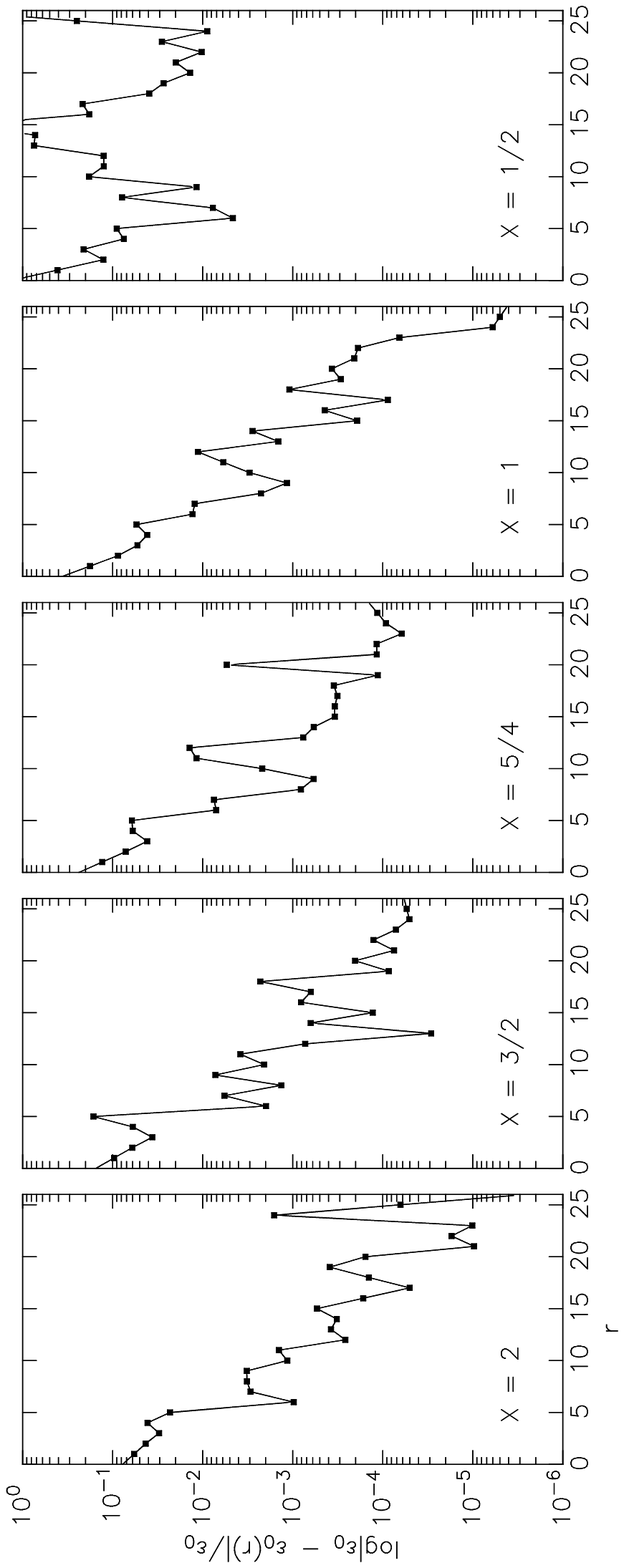}
 \caption
   [The relative error of the ground state energy evaluated by a 
    $ \hbox{[}r/r\hbox{]} $ diagonal Pad\'e approximation to $ E(t) $
    versus the truncation order $ r $ for various couplings $ x $.]{}
\label{fig-texp-gse}
\end{figure}

In the first study we employ only the simplest extrapolation strategy for
the $ t $-expansion, namely diagonal Pad\'e approximation. 
In Figure~(\ref{fig-texp-gse}) we have plotted the relative error 
of the ground state energy evaluated by a $ [r/r] $ diagonal Pad\'e
approximation versus the truncation order $ r $ for various couplings
$ x $. Clearly the error is not smooth or strictly monotonic with
$ r $, as the Pad\'e Approximant can change significantly from one order
to the next but the overall trends are obvious. The general convergence
is quite good, decreasing exponentially rapidly with order, with
the rate of decrease highest for large $ x $. As $ x $ approaches unity
this rate of convergence drops, but doesn't vanish altogether, so
we have the remarkable result that even at the critical point we
still have substantial convergence. For couplings less than unity,
such as the example given, we do not see any convergence beyond a 
minimal value.

\begin{figure}[htb]
 \vskip 22.7cm
 \includegraphics{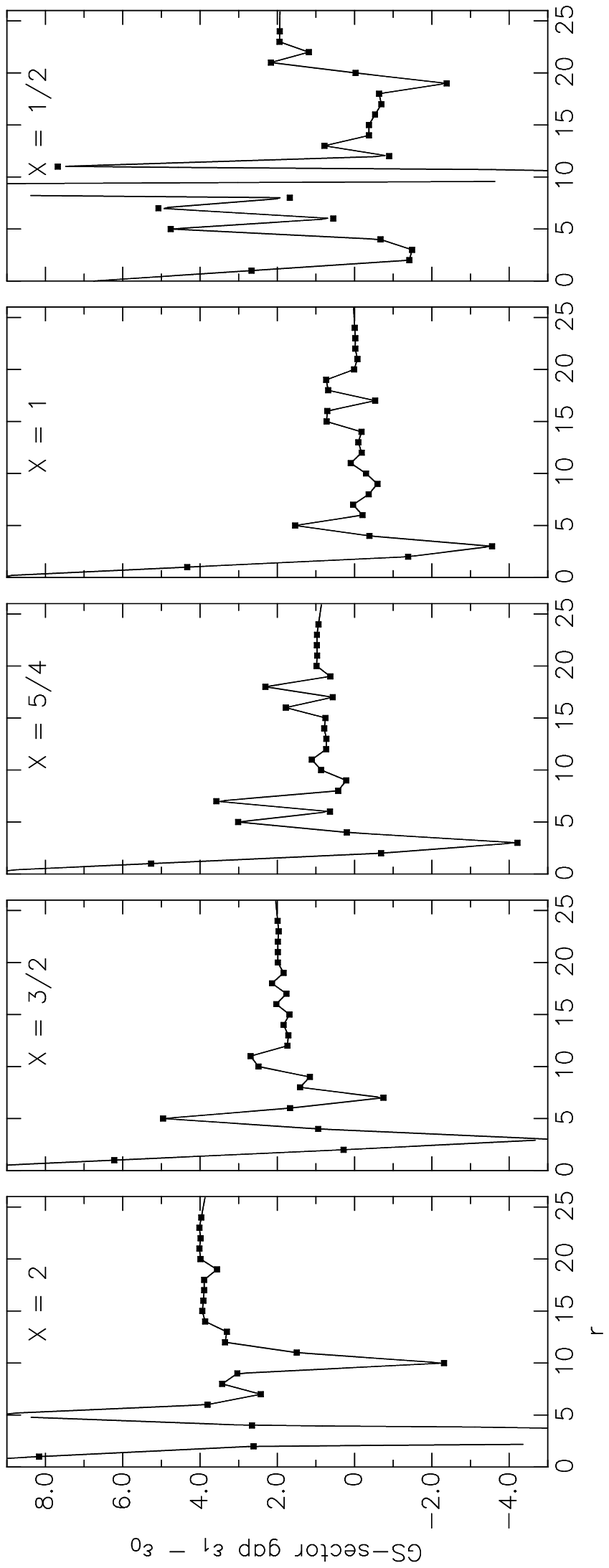}
 \caption
   [The excited state gap evaluated by a $ \hbox{[}r/r\hbox{]} $
    diagonal Pad\'e approximation to $ -E''(t)/E'(t) $ versus the 
    truncation order $ r $ for various couplings $ x $.]{}
\label{fig-texp-gap}
\end{figure}

For the estimates of the excited state gap we expanded the derivative
ratio, given in Eq.~(\ref{HW-thm}), as a Taylor series then formed the
diagonal $ [r/r] $ Pad\'e Approximations to this and made the 
extrapolation. In Figure~(\ref{fig-texp-gap}) we have plotted this gap 
versus the truncation order $ r $ for the various couplings.
Again for small orders there are wild fluctuations as the Pad\'e
Approximant shifts from one order to the next but these die out and
converge quickly to the correct values, given by Eq~(\ref{itf-gap}).
And again, most remarkably, the gaps converge as the coupling tends
to unity, even at the transition point, and there there is evidence
of a late convergence even in the other phase, at $ x=1/2 $.

\begin{figure}[htb]
 \vskip 22.7cm
 \includegraphics{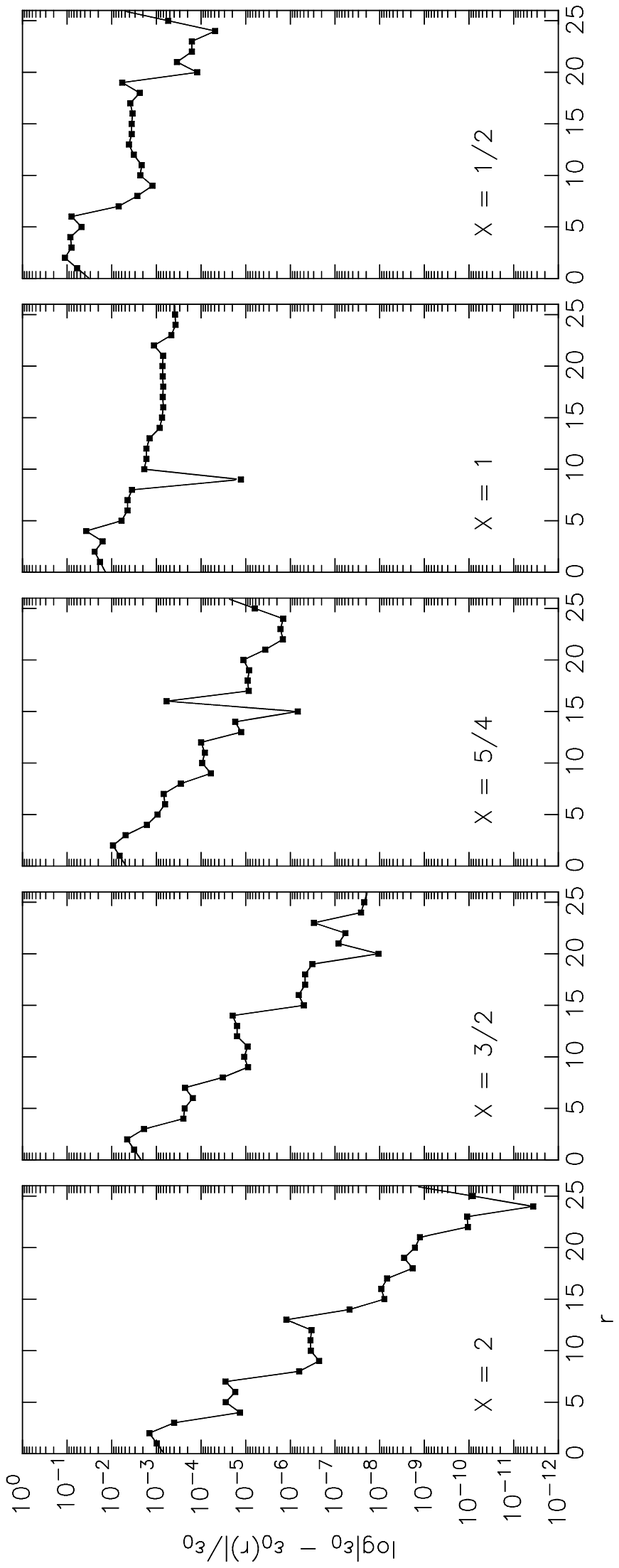}
 \caption
   [The ground state energy evaluated with CMX-LT method using
    a $ r\times r $ matrix versus the truncation order $ r $ 
    for various couplings $ x $.]{}
\label{fig-cmxlt-gse}
\end{figure}

In the CMX studies we first consider the CMX-LT variant.
In Figure~(\ref{fig-cmxlt-gse}) we have plotted the relative error of
the expression, Eq.~(\ref{cmxlt-gse}), truncated to $ r\times r $ matrix
versus the truncation order $ r $ for the same sequence of 
couplings $ x $. While there are still fluctuations in going from
order to order these are less than for the $ t $-expansion, and the
rate of convergence is much faster. The slowing down of the rate
of convergence as the coupling approaches the critical one is also
clearly evident, although even at the transition point there is
still a respectable level of convergence. In the other phase, at
$ x=1/2 $ there is some improvement, even if the fluctuations are
much larger, and the error saturates again at a modest value.
For the HW variant of the CMX, the ground state energy estimate failed
to converge to the correct value even at $ x=2 $, rather it converged
to another value with a relative error of $ 0.027\% $. In the case 
of the CMX-SD method it actually diverged for this coupling value,
showing no convergent tendencies within the first $ 20 $ cumulants.

\eject

\section{Conclusions}

In this work we have found the exact generating function and cumulants
for the Ising Model in a transverse field for arbitrary coupling constant,
with respect to a simple trial state taken to be the exact eigenstate at 
infinite coupling. For couplings larger than the critical coupling we
have found that the Horn-Weinstein function is analytic in the right
half of the complex plane, and has cuts along an infinite number of 
curves lying in the left half of the complex plane. When the coupling 
strength is less than the critical one, these singular curves cross both
halves of the complex plane and the function is only analytic in a strip
along the real axis. The radius of convergence of the Taylor expansion
for the H-W function is finite, but does disappear for large coupling,
however this is not quantitatively important for the $ t $-expansion or
Connected Moments Expansion. The Asymptotic form of the H-W function
for large real positive argument displays a crossover as the coupling
strength goes through the critical point. For couplings larger than the
critical value one has a dominant exponential decay governed by the 
excited state gap within the ground state sector, along with an algebraic
series of non-integral powers. At the critical point there is pure
algebraic decay with integral exponents. And for couplings less than the
critical one there is exponential decay determined by the first excited
state gap. Results of the convergence for the $ t $-expansion reveal that
with the simplest extrapolation method, the ground state and excited state
gap estimates converge in a systematic way for all couplings larger than
the critical one, and even at the critical point, although with limited 
accuracy there. 
Only the CMX-LT variant converged, of the CMX methods, and in this case
the ground state energies converged quite rapidly for all couplings in 
the above range, again even at the critical point.
Because the actual character of the H-W function in these Models is 
rather different from that commonly assumed, and that these Models are not
atypical of other strongly coupled Models, then extrapolation strategies
in the $ t $-expansion that employed this qualitative knowledge would be 
expected to have better convergence performance. Furthermore these
examples provide concrete forms for new variants of the Connected Moment
Expansion to be developed.

\eject

\bibliographystyle{amsplain}
\bibliography{moment,texp,cmx,itf,alm,math}
\vspace*{30mm}
\begin{center}
\bf Acknowledgements
\end{center}
This work was supported by the Australian Research Council.
One of us (NSW) would like to acknowledge the hospitality of the
Institute of Mathematical Sciences, Chennai India, where this
work was started and also numerous helpful discussions with L.C.L. Hollenberg,
C.J. Hamer and the members of the Condensed Matter Theory Group at the
University of NSW.
\vfill\eject

\listoffigures
\clearpage
\eject

\eject

\eject

\eject

\eject

\end{document}